\documentclass[]{aa}
\usepackage{graphicx}
\usepackage{textcomp}
\usepackage{gensymb}
\usepackage[varg]{txfonts}
\usepackage{color}
\usepackage{minipage}
\usepackage{appendix}
\usepackage{multirow}
\usepackage{amsmath}
\usepackage{natbib}
\usepackage[colorlinks,linkcolor=blue,citecolor=blue,linktocpage=true,breaklinks,plainpages=false,urlcolor=blue]{hyperref}

\definecolor{DarkGreen}{rgb}{0,0.8,0}

\bibpunct{(}{)}{;}{a}{}{,}
\begin{document}
\title{The non-LTE formation of the \ion{Fe}{I} 6173\,\AA{} line in the solar atmosphere}
\author{H. N. Smitha \inst{1}, M. van Noort \inst{1}, S. K. Solanki \inst{1,2}, J.~S. {Castellanos~Dur\'{a}n} \inst{1}}
\institute{Max-Planck-Institut f\"ur  Sonnensystemforschung, 
Justus-von-Liebig-Weg 3, 37077 G\"ottingen, Germany \and
School of Space Research, Kyung Hee University, Yongin, Gyeonggi, 446-701, Republic of Korea\\
\email{smitha@mps.mpg.de}}
\titlerunning{The non-LTE formation of the \ion{Fe}{I} 6173\,\AA{}}
\authorrunning{Smitha et al.}

\abstract
{The current analysis is dedicated to a detailed investigation of the non-Local Thermodynamic Equilibrium (NLTE) effects influencing the formation of the  \ion{Fe}{I} 6173\,\AA{} line, which is widely used by many instruments including the Helioseismic and Magnetic Imager (HMI) on-board the Solar Dynamics Observatory (SDO) and the Polarimetric and Helioseismic Imager on board the Solar Orbiter.
We synthesize the Stokes profiles in a snapshot of a three dimensional magnetohydrodynamic simulation of the solar photosphere under both LTE and NLTE conditions. The simulation cube contains a sunspot and a plage region around it. The LTE and NLTE Stokes profiles formed in different features are compared and analysed.
NLTE effects are evident in both intensity and polarization profiles. For the 6173\,\AA{} line, UV overionization is the dominant NLTE mechanism, and scattering effects are much less important. {In addition to Fe, an NLTE treatment of Si, Mg and Al is necessary to set the right photon density in the UV. This is found to further enhance the LTE departures compared to the case where Fe alone is treated in NLTE.} These effects in the Stokes profiles survive even when the profiles are averaged spatially or sampled on a coarse wavelength grid such as that used by the SDO/HMI and other magnetographs.
The deviations from the LTE profiles are stronger in the \ion{Fe}{I} 6173\,\AA{} compared to the 6301\,\AA--6302\,\AA{} lines because in case of the latter, line scattering compensates the effect of UV overionization. Based on the nature of departures from LTE, treating the 6173\,\AA{} line in LTE will likely result in an over-estimation of temperature and an under-estimation of the magnetic field strength.}

\maketitle

\section{Introduction}
\label{sec:intro}
Iron lines crowd the solar spectrum from ultraviolet (UV) to infrared (IR) wavelengths. Lines of neutral iron are among the most widely used diagnostic lines for the solar photosphere and in particular for photospheric magnetism \citep{1982A&A...115..104R, 1988A&A...189..243S, 1988ASSL..138..185R}. The photospheric iron lines are routinely modeled assuming Local Thermodynamic Equilibrium (LTE) conditions.

\citet{1972PhDT.........7L} and \citet{1972ApJ...176..809A} were among the first to discover the departure from LTE in \ion{Fe}{I} lines due to imbalances in radiative bound-free transitions. UV photons with energies that match the bound-free jumps of the well populated atomic levels overionize these levels.  Although the overionization mainly occurs from higher levels with energies around $4$\, eV, its effects get shared over all atomic levels due to strong collisional and radiative coupling \citep{1988ASSL..138..185R}.  As a result, the atomic level populations are no longer given by Saha-Boltzmann statistics. The signatures of this non-LTE (NLTE) effect are imprinted in all iron lines observed in the UV, visible and infrared wavelength regions.
UV overionization is not just limited to iron atoms but affects all minority species including alkalis \citep{1982A&A...115..104R, 1992A&A...265..237B, 2021arXiv210302369R}. In addition to UV overionization, the spectral lines can also be affected by other NLTE mechanisms such as the resonance scattering which is specific to  particular spectral lines \citep{1992A&A...265..237B}. The observed spectral profiles are shaped by a combination of different NLTE mechanisms. Depending on the atmospheric conditions and the properties of the individual line, the resulting profile can have a shape close to its LTE equivalent, or it can be significantly different. 

A number of studies have investigated the role of  NLTE effects in the formation of the famous \ion{Fe}{I} 6301\,\AA{} -- 6302\,\AA{}  line pair (also refereed to as the $6300\,\AA$ pair) and their influence on the inferred atmospheric models \citep[see][]{1982A&A...115..104R, 2001ApJ...550..970S, 2012A&A...547A..46H, 2013A&A...558A..20H, 2015A&A...582A.101H, 2020A&A...633A...157S, 2021A&A...647A..46S}. In the present study, we investigate the formation of \ion{Fe}{I} 6173\,\AA{}, another widely utilized line. This line has been extensively used in inferring solar properties, because it is a clean line without any blends, which makes it advantageous for velocity measurements, e.g. for helioseismic applications, and because it is a normal Zeeman triplet ($J_l=1$, $J_u=0$) with a large Land{\'e} factor ($\mathrm{g_{eff}}=2.5$), which makes it favourable for measuring magnetic fields. It is used by the Helioseismic and Magnetic Imager \citep[HMI,][]{2012SoPh..275..207S} onboard the Solar Dynamics Observatory \citep[SDO, ][]{2012SoPh..275....3P} to capture full-disk images of the Sun nearly continuously. It is also used by the Polarimetric and Helioseismic Imager on the recently launched Solar Orbiter \citep[SO/PHI][]{2020A&A...642A..11S, 2020A&A...642A...1M}. The Photospheric Magnetic Field Imager (PMI) on ESA's Vigil mission to the Lagrange point L5 is also designed to observe the Sun in this line \citep{2020JSWSC..10...54S}.  

Ground based observatories such as the Swedish Solar Telescope \citep[SST,][]{2003SPIE.4853..341S}, and the Daniel K. Inouye Solar Telescope \citep[DKIST,][]{visp} also observe this line. However, the modeling of the 6173\,\AA{} line is often done under the assumption of LTE \citep[for example,][]{2009A&A...494.1091B, 2017A&A...608A..97F, 2020A&A...635A.210P, 2021A&A...649A.106Y} or by using the even simpler Milne-Eddington approximation \citep[for example][]{2011SoPh..273..267B, 2014SoPh..289.3531C, 2020A&A...641L...5J, 2022A&A...662A..88V, 2022arXiv220308172J}. {Previously, the 6173\,\AA{} line has been treated in NLTE in only a few papers including \citet{2020ApJ...893...24S}, \citet{2021ApJ...915...16M}, \citet{2022A&A...661A..59D}, and also by \citet{2013ApJ...778..175H}, \citet{2018ApJ...857L...2H} for flare diagnostics.}

{The present study is dedicated to understanding the different sources of NLTE effects influencing the formation of the \ion{Fe}{I} 6173\,\AA{} line. We find that these effects are prominently seen in the Stokes profiles of this line. Interestingly, they even survive when the profiles are spatially averaged or spectrally sampled over a coarse grid. } In fact the deviations from LTE in this line can be stronger than for the \ion{Fe}{I} 6301\,\AA -- 6302\,\AA{} line pair. The details are presented in the sections below. 

\section{MHD cube and profile synthesis}
\label{sec:cube}
To investigate the NLTE effects in the \ion{Fe}{I} 6173\,\AA{} line, we computed its Stokes profiles from a three dimensional radiation magneto-hydrodynamic (MHD) cube. We choose a high resolution non-grey sunspot cube by \citet{2012ApJ...750...62R}\footnote{\protect\url{http://download.hao.ucar.edu/pub/rempel/sunspot_models/Fine_Structure}} generated using the MURaM code \citep{2005A&A...429..335V}. The cube has a grid spacing of $12$\,km in the horizontal direction, and $8$\,km in the vertical direction. 

\begin{figure}
    \centering
    \includegraphics[width=0.5\textwidth]{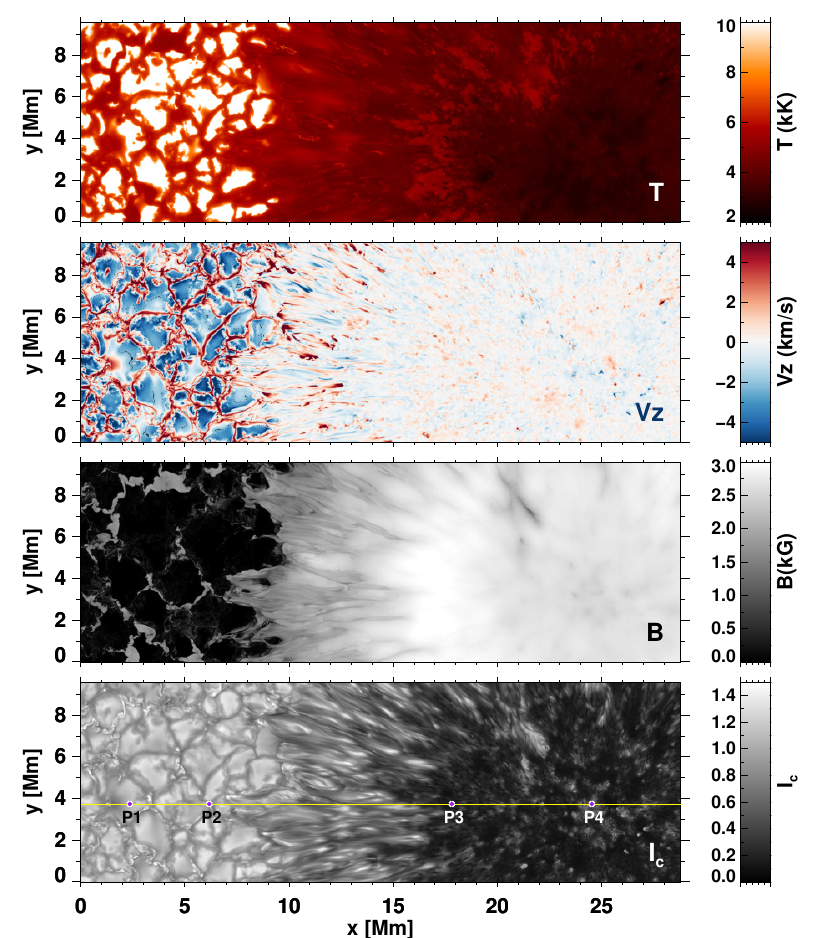}
     \caption{Maps of temperature (first panel), line-of-sight velocity (second panel) and magnetic field strength (third panel) over the chosen region in the MHD cube at a geometric height $z=0$\,km. Note that positive velocities denote downflows. 
     Continuum intensity image at 6170\,\AA{} is shown in the bottom panel. Here the points P1, P2, P3 and P4 are the four pixels at which we later present a detailed comparison between the Stokes profiles computed in LTE and NLTE.}
    \label{fig:cont}
\end{figure}

To save computing time, we synthesized the profiles over a small patch of the cube, covering the sunspot and the plage region around it. The temperature, vertical velocity and magnetic field strength maps at $z=0$\,km are plotted in the first three panels of Figure~\ref{fig:cont}. Here $z=0$\,km corresponds to the layer where $\log({\tau_{5000}})$ in the part of the simulation box outside the sunspot is zero, on average. The Stokes profiles were computed at the heliocentric angle $\mu=1$ in LTE and NLTE along every column using the RH 1.5D code \citep{2001ApJ...557..389U, 2015A&A...574A...3P}. {No additional microturbulent velocity was used to broaden the line profiles.} The continuum intensity image of this patch at 6170\,\AA{} is shown in the bottom panel of Figure~\ref{fig:cont}. The four pixels, P1 to P4, marked in this figure, are sample pixels at which we later discuss the Stokes profiles in detail. 

According to \citet{2021arXiv210302369R}, a proper NLTE treatment of the iron lines demands a large Fe atom model, {treatment of Si, Al and Mg in NLTE to set the radiation field in the UV}, and the UV line haze. For iron, we here use a 33 level atom model which is a part of the RH 1D code\footnote{\url{https://github.com/han-uitenbroek/RH/blob/master/Atoms/Fe33.atom}}\citep{2001ApJ...557..389U}. In this model, 31 levels belong to \ion{Fe}{I}. The other two levels are the ground states of \ion{Fe}{II} and \ion{Fe}{III}, respectively. The atomic levels are coupled by 63 bound-bound and 31 bound-free transitions. The atom includes photo-ionization cross-sections of the bound-free transitions computed using the hydrogenic approximation. 

{Metals such as Si, Mg, Fe and Al have ionization edges in the UV. They play an important role in setting the level of angle averaged mean intensity $J_{\nu}$ in the UV. Additionally, they are also the most abundant electron-donor atoms in the solar temperature minimum  \citep{1981ApJS...45..635V, 2003rtsa.book.....R,  2021arXiv210302369R}. The electrons contributed by them affect the formation of \ion{H}{$^-$}. For a complete NLTE treatment, it is important to treat all these metals in NLTE and also self-consistently compute the electron densities in NLTE. However, in the present study we restricted ourselves to treating the former effect only. That is, the electron densities are computed by assuming LTE ionization. Additionally, for the major part of the cube, we treat only Fe in NLTE to save computing time. Over a small stripe of the cube, we compute the Stokes profiles by treating also the other metals (Si, Mg and Al) explicitly in NLTE. A detailed discussion on this is presented in Section~\ref{sec:detailed_comp_ed}. }

The effects of the UV line haze are incorporated using opacity fudging described in \citet{1992A&A...265..237B}. The fudge factors include frequency-dependent multiplication factors to the total \ion{H}{$^-$} opacity for $\lambda > 210$\,nm and at shorter wavelengths, the metal opacity is enhanced. These factors were derived by \citet{1992A&A...265..237B} by fitting the continuum computed from the VAL3C model atmosphere \citep{1981ApJS...45..635V} with the observed continuum.

\begin{figure*}
    \centering
    \includegraphics[width=0.80\textwidth]{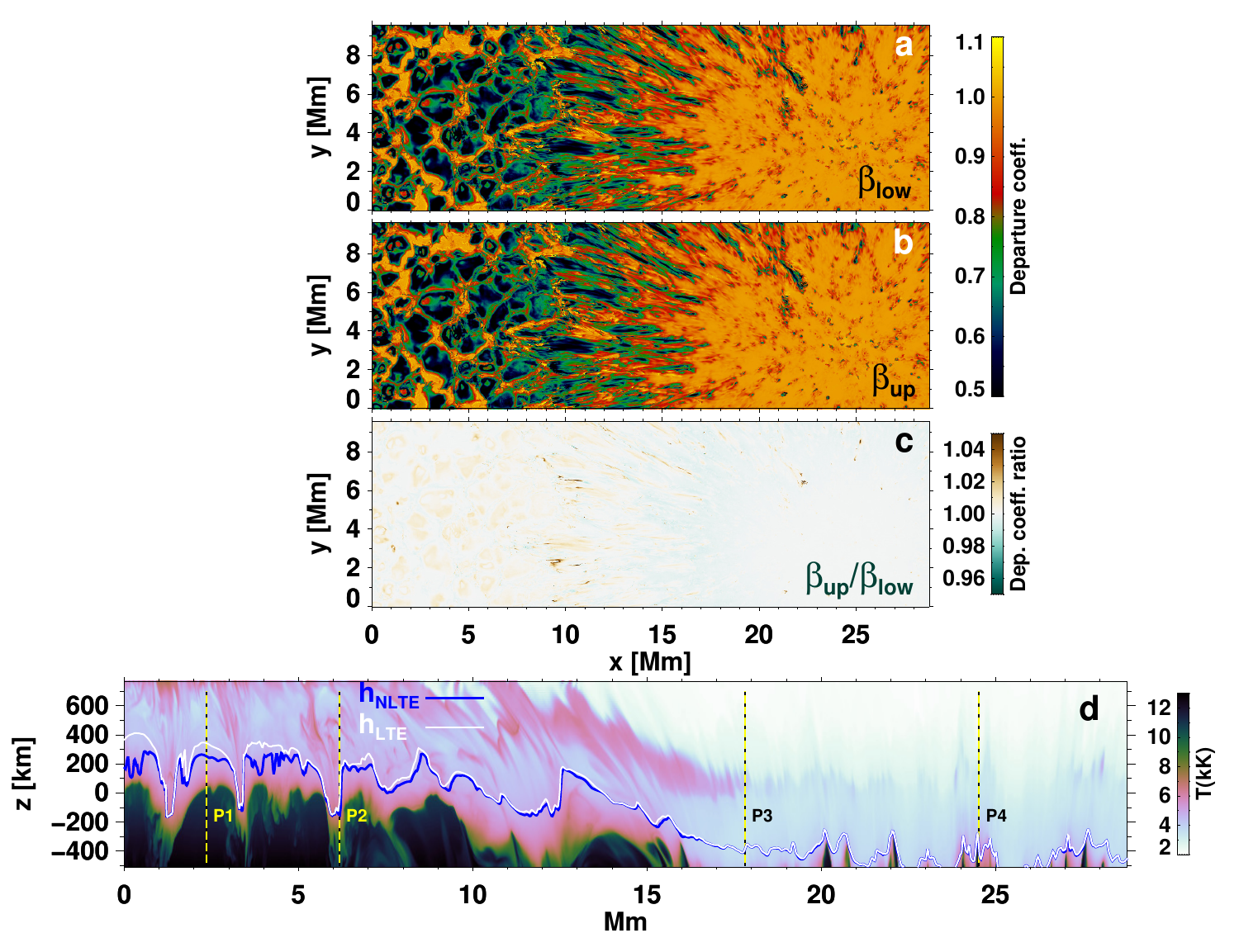}
    \caption{Maps of departure coefficients at heights where   $\tau_{\lambda_0}=1$, with $\lambda_0$ being the rest wavelength of the \ion{Fe}{I} 6173\,\AA{} line. Top panel: departure coefficients of the lower level ($\beta_{low}$); middle panel: departure coefficient of the upper level ($\beta_{up}$); bottom panel: ratio $\beta_{up}/\beta_{low}$. Heights corresponding to  $\tau_{\lambda_0} = 1$ surface computed from LTE (yellow line) and NLTE (blue line) are plotted in panel d. In the background is the temperature map of the slice marked by the horizontal yellow line in Figure~\ref{fig:cont}. {The positions of the four sample pixels chosen for detailed analysis are marked by vertical lines in panel d.}}
    \label{fig:dep-coeff}
\end{figure*}

\label{sec:dep_coeff}
\begin{figure*}
    \centering
    \includegraphics[width=0.50\textwidth]{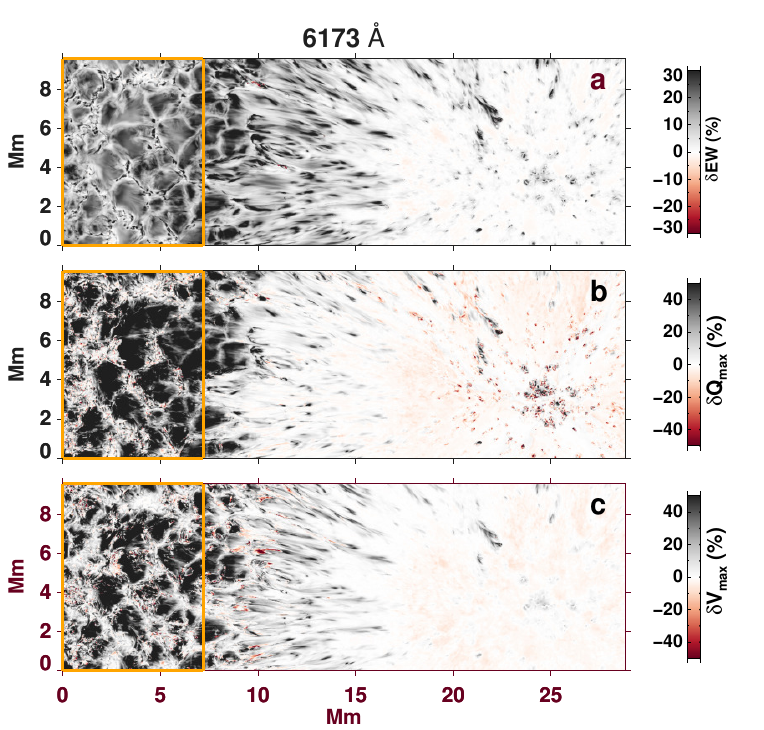}
    \includegraphics[width=0.28\textwidth]{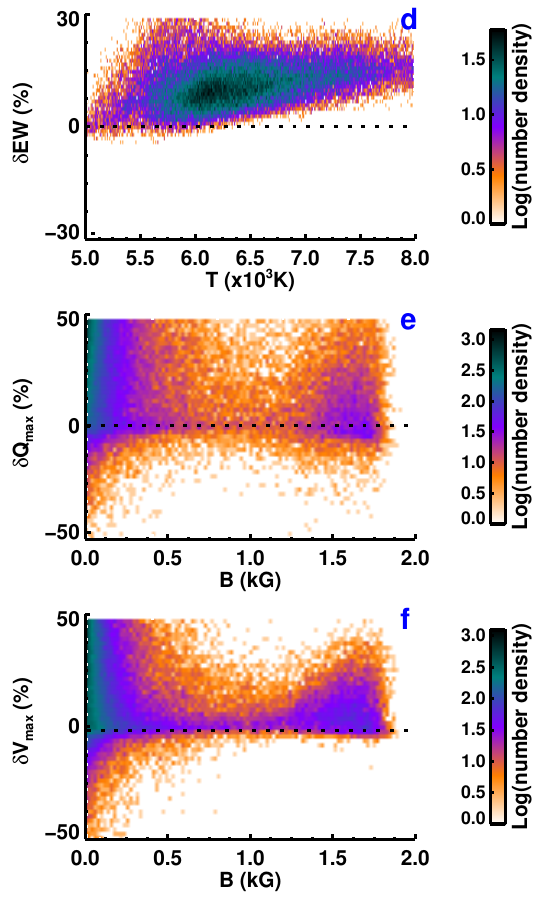}
    \includegraphics[width=0.18\textwidth]{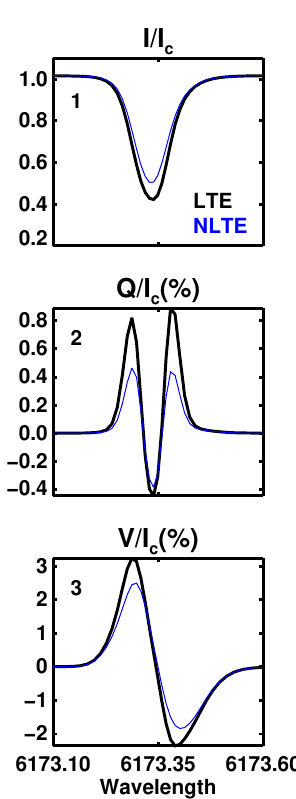}\\
    \includegraphics[width=0.32\textwidth]{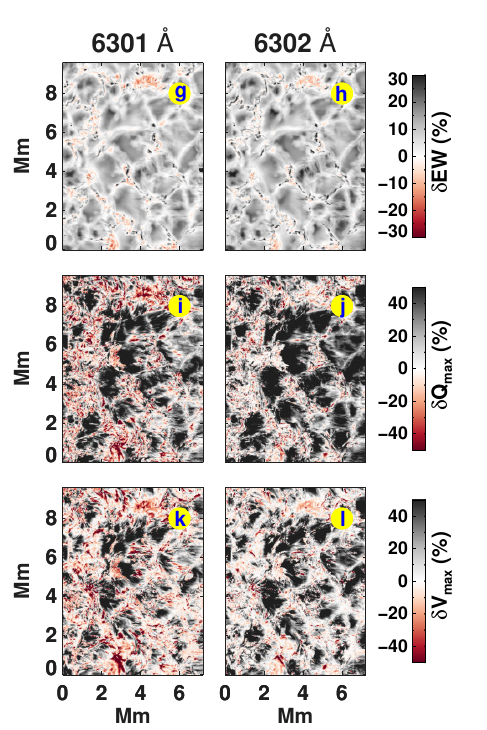}
    \includegraphics[width=0.28\textwidth]{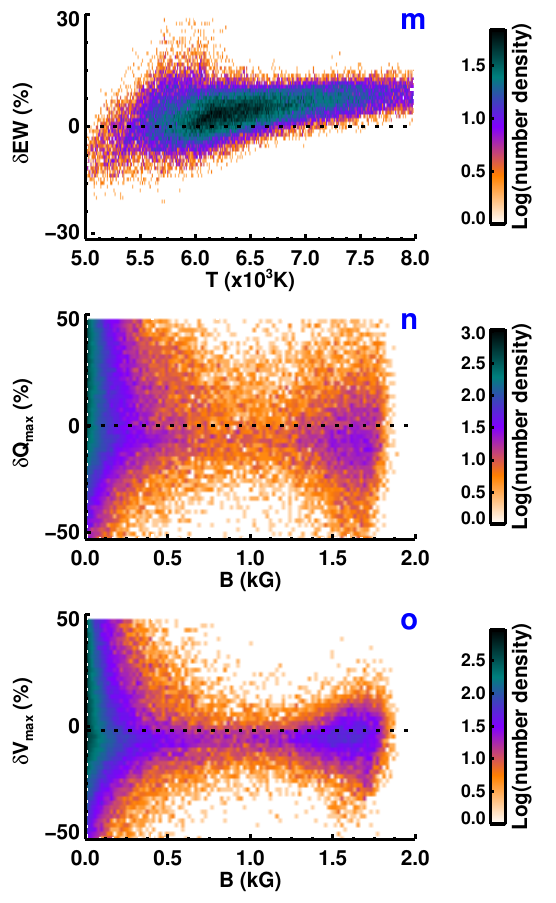}
    \includegraphics[width=0.18\textwidth]{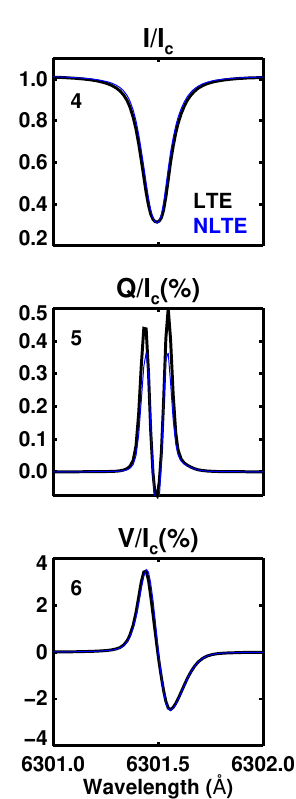}
    \caption{Comparison between the intensity, linear polarization and circular polarization profiles computed in LTE and NLTE. Panels a - c show maps of relative differences $\delta EW$,  $\delta Q_{\rm max}$ and $\delta V_{\rm max}$, respectively, for the \ion{Fe}{I} 6173\,\AA{} line (these quantities are defined in Equation~\ref{eq:rel_diff}). Panels g-l show the same for the 6301\,\AA{} -- 6302\,\AA{} line pair, but over a smaller region outside the sunspot indicated by the orange boxes in paenels a-c. Panels d-f display  scatter plots between various quantities for pixels from within the orange box. They demonstrate the variation in $\delta EW$ as a function of temperature ($T$ at $z=0$\,km), and the variation in $\delta Q_{\rm max}$ and $\delta V_{\rm max}$ as a function of the magnetic field strength $B$ at $z=0$\,km. Panels m-o show similar scatter plots for the 6301\,\AA line. The stokes profiles of the 6173\,\AA{} line spatially averaged over the orange box are plotted in panels 1-3, those of the similarly averaged 6301\,\AA{} line are plotted in panels 4-6.}
    \label{fig:ew}
\end{figure*}

\section{NLTE mechanisms}
\label{sec:nlte_mech}
According to \cite{1992A&A...265..237B}, five important NLTE mechanisms affect minority species. They are: resonance line scattering and photon loss, photon suction, UV overionization, infrared over recombination, and UV line pumping. For iron, UV overionization and resonance scattering are the two dominant mechanisms \citep{2001ApJ...550..970S}. We briefly describe them below. More details can be found in \citet{1982A&A...115..104R, 1988ASSL..138..185R, 1992A&A...265..237B, 1992A&A...265..257B, 2021arXiv210302369R}.

\textit{UV overionization:} The radiation temperature at wavelengths between 200 nm and 500 nm exceed the electron temperature, $T_e$ \citep[][]{1985cdm..proc...67A, 1988ASSL..138..185R, 1992A&A...265..237B}. Hence the UV photons are superthermal with angle-averaged mean intensity $J_{\nu}$ exceeding the Planck function $B_{\nu}(T_e)$. This excess UV radiation incident onto the photosphere from the cool temperature minimum region is absorbed by the minority species including iron, which have well populated atomic levels with energies between $2$\,eV - $4$\,eV from the continuum, resulting in UV overionization.  Excess ionization leads to line opacity deficits in \ion{Fe}{I}, and an opacity excess in \ion{Fe}{II}.

\textit{Resonance scattering in lines:} The second dominant NLTE mechanism affecting iron lines is resonance scattering in atomic bound-bound transitions. This is a property of the individual lines. According to \citet{2001ApJ...550..970S}, for iron lines with intermediate to high excitation potential ($> 2$\,eV) the source function deviates from the Planck function and when the excitation potential is $< 2$\,eV, the source function follows the Planck function \citep[see also][]{1988ASSL..138..185R, 1988A&A...189..243S}. A source function deficit results from the upper level departure coefficient being less than that of the lower-level. This is due to outward photon-losses. Scattering propagates this to layers below the surface where the optical depth at line center is greater than unity \citep{1992A&A...265..237B, 1994A&A...288..860C}. Source function deficit makes the NLTE line stronger than the LTE line.

The line opacity deficit due to UV overionization and the source function deficit due to scattering affect the spectral profiles in opposite ways. While the former makes the spectral line shallower, the latter makes the line stronger and deeper than the LTE line. For the iron line pair at 6301\,\AA{} -- 6302\,\AA{}, with an excitation potential of 3.6 eV, both these effects play a role \citep{2020A&A...633A...157S}. Depending on the atmospheric conditions and the dominant NLTE mechanism, the profile computed in NLTE can either be stronger or weaker than the LTE line. In some cases when both the effects are equally strong, the NLTE line can be identical to the LTE line.

The \ion{Fe}{I} 6173\,\AA{} line has an excitation potential of $2.2$\,eV. Its radiative width (Einstein $A$ coefficient) is an order of magnitude smaller than that of the 6301\,\AA{} line. The collisional coupling is stronger and the photon destruction probability is much higher. Further, the line interlocking effects forces the source function to follow the Planck function by setting the upper-level departure coefficient equal to that of the lower-level. Therefore, the 6173\,\AA{} line is not as much affected by scattering as the 6300\,\AA{} line pair, and UV overionization is the only dominant NLTE mechanism affecting it.

\subsection{Departure coefficients}
\label{subsec:dep_coeff}
 The departure coefficient, $\beta$,  is defined as the deviation of an atomic level population from its LTE value \citep{1972SoPh...23..265W}
\begin{equation}
\beta = \frac{n_{i}}{n_{i}^*},    
    \end{equation}
where $n_{i}$ and $n_{i}^{*}$ are the NLTE and LTE populations of level $i$, respectively. 

 The line opacity deficit from the excess UV ionization scales with the lower level departure coefficient, $\beta_{\rm low}$. It changes the height where the line optical depth reaches unity, that is the $\tau({\lambda_0})=1$ surface with $\lambda_0$ being the rest wavelength of the line. We denote this height as $h_{\rm LTE}$ and $h_{\rm NLTE}$ for LTE and NLTE computations, respectively. In the rest of this paper, $h_{\rm LTE}$ and $h_{\rm NLTE}$ are used as reference heights to compare different physical quantities, and are sometimes referred to as the height of formation of the line. However, we are aware of the limitations and uncertainties introduced by such a definition \citep{1996A&A...314..295S}. It is used here only for simplicity. More discussion on this can be found in Section~\ref{sec:detailed_comp}.
 
  A map of the lower-level departure coefficient $\beta_{\rm low}$ at $h_{\rm NLTE}$ over the region of interest for the 6173\,\AA{} line is shown in the top panel of Figure~\ref{fig:dep-coeff}. The $\beta_{\rm low}$ at $h_{\rm NLTE}$ is less than one in granules, intergranular lanes, penumbral filaments, and umbral dots. This is due to UV overionization. Over hotter granules, $\beta_{\rm low}$ is lower than over cool intergranular lanes. This is because, the UV radiation is more superthermal ($J_{\nu} > B_{\nu}$) over granules  than over intergranular lanes \citep{1992A&A...265..257B}. The $\beta_{\rm low}$ at $h_{\rm NLTE}$ is close to unity in the umbra and in regions of strong magnetic field outside the sunspot, seen by comparing Figure~\ref{fig:dep-coeff} to the third panel of Figure~\ref{fig:cont}. Even though $\beta_{\rm low}$ is close to one in bright magnetic structures, later in Sections~\ref{subsec:int} and \ref{sec:detailed_comp}, we discuss how the LTE departures in the Stokes profiles can still be quite strong. 
 
  The line opacity deficit translates into $h_{\rm NLTE} < h_{\rm LTE}$ \citep{2001ApJ...550..970S}. Panel d of Figure~\ref{fig:dep-coeff} shows $h_{\rm NLTE}$ and $h_{\rm LTE}$, overplotted over a vertical slice through the atmosphere, marked by the yellow line on the continuum image in Figure~\ref{fig:cont}. Outside the sunspot, $h_{\rm NLTE}$ is lower than $h_{\rm LTE}$, while in the sunspot $h_{\rm NLTE}$ is close to $h_{\rm LTE}$. 
 
 The deviations in the line source function $S^{l}_{\nu}/B_\nu$ due to bound-bound scattering is proportional to $\beta_{\rm up}/\beta_{\rm low}$ \citep{1992A&A...265..237B}. Panels b and c in Figure~\ref{fig:dep-coeff} show maps of $\beta_{\rm up}$ and $\beta_{\rm up}/\beta_{\rm low}$, respectively. Almost everywhere within the chosen region of the cube,  $\beta_{\rm up}$ follows $\beta_{\rm low}$ and the ratio $\beta_{\rm up}/\beta_{\rm low}$ is close to one. The line source function follows the Planck function, indicating the weaker role of resonance scattering for the 6173\,\AA{} line. In panel c, we see a few patches where $\beta_{\rm up}/\beta_{\rm low}$ is slightly greater than one. In these pixels, the MHD cube has strong gradients in temperature, and the steep increase in $T$ around $h_{\rm NLTE}$ gives rise to $\beta_{\rm up}/\beta_{\rm low} > 1$. This increase is localised and occurs only over a narrow range of heights in the cube. It does not have an impact on the overall line formation. 

\section{Effects on Stokes profiles}
\label{sec:Stokes_lte_nlte}
The NLTE effects alter the equivalent widths ($EW$) and depths of the intensity profiles  \citep[][]{1982A&A...115..104R, 1988A&A...189..243S, 2001ApJ...550..970S, 2015A&A...582A.101H}, and also the strength of the polarization profiles \citep{2020A&A...633A...157S}. In Figure~\ref{fig:ew}, we compare the LTE and NLTE profiles using their $EW$, and $Q/I_c$, $U/I_c$ or $V/I_c$ peak amplitudes, denoted as $Q_{\rm max}$, $U_{\rm max}$ or $V_{\rm max}$, respectively. Relative changes in these quantities due to the NLTE effects are given by
\begin{equation}
    \delta x =\frac{x^{\rm LTE} - x^{\rm NLTE}}{x^{\rm LTE}},
    \label{eq:rel_diff}
\end{equation}
where $x$ can be $EW$, $Q_{\rm max}$, $U_{\rm max}$ or $V_{\rm max}$. A positive value of $\delta x$ indicates line weakening due to overionization while a negative value indicates line strengthening mainly due to resonance scattering.

In the same figure, we also show maps of the corresponding quantities for the 6301\,\AA{} -- 6302\,\AA{} line pair. For this comparison, we have selected a small region, marked by the orange rectangle over which the Stokes profiles of this line pair were computed. This region was chosen outside the sunspot because the 6173\,\AA{} line is not affected by NLTE effects in the umbra (see Figure~\ref{fig:dep-coeff}). The details are discussed in the sections below.

\begin{figure*}
    \centering
    \includegraphics[width=0.95\textwidth]{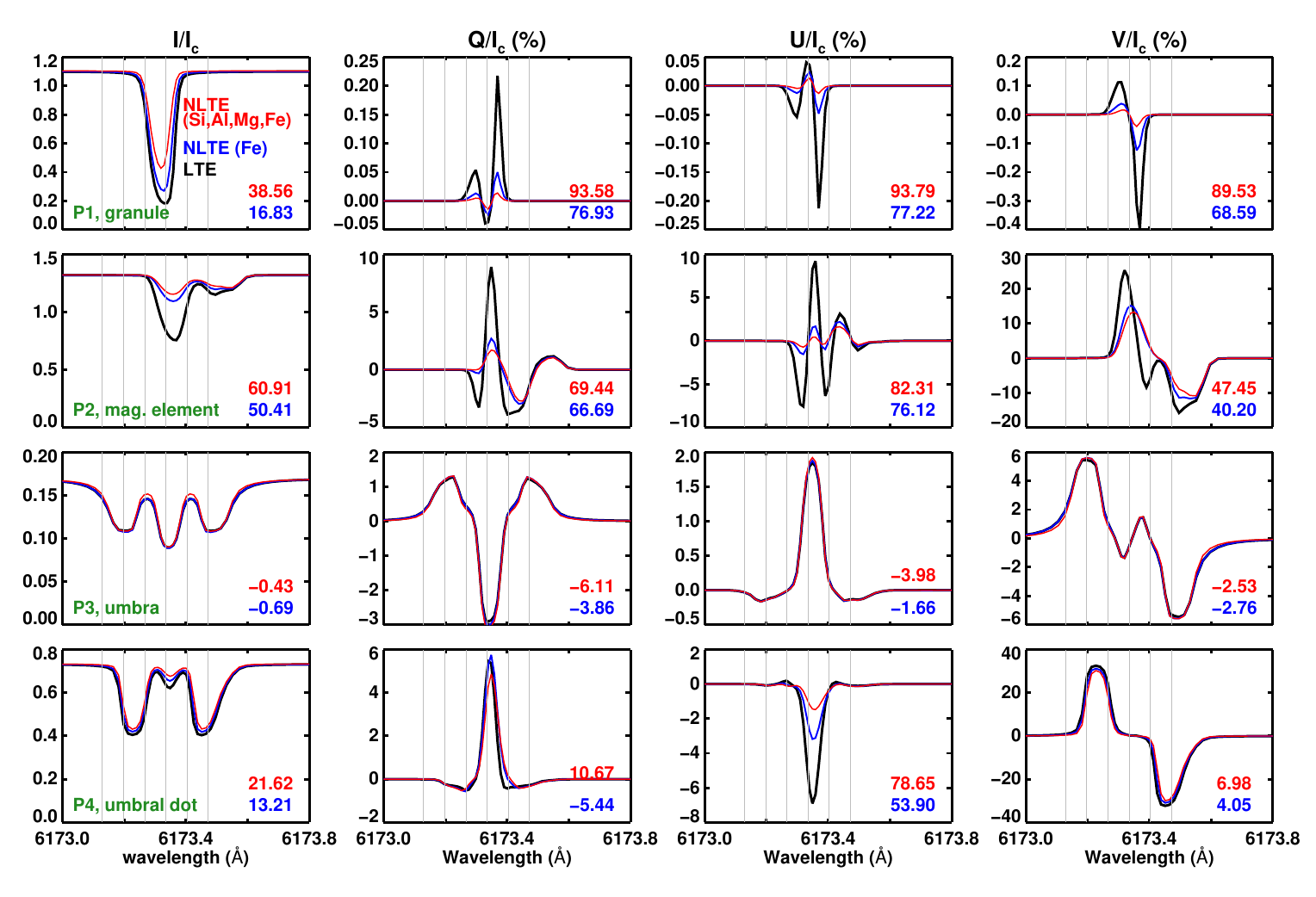}
      \caption{Comparison between the Stokes profiles computed in LTE and NLTE at the four sample pixels P1, P2, P3 and P4 indicated in Figure~\ref{fig:cont}. The black curves are computed in LTE. The blue curves are computed by treating only iron in NLTE. The red curves are computed by treating Fe as well as {Si, Al, and Mg} in NLTE. The four columns show Stokes $I/I_c$, $Q/I_c$, $U/I_c$ and $V/I_c$, respectively, expressed as percentage. $I_c$ is the spatially averaged continuum intensity over the non-spot region of the cube. The six vertical lines mark the positions of the grid points sampled by HMI onto which we later interpolate the Stokes profiles. The numbers in the bottom right corner of each panel are the $\delta EW$, $\delta Q_{\rm max}$, $\delta U_{\rm max}$ and $\delta V_{\rm max}$ respectively, for the profiles in red and blue.}
    \label{fig:stokes_lte_nlte}
\end{figure*}

\begin{figure*}
    \centering
    \includegraphics[width=0.95\textwidth]{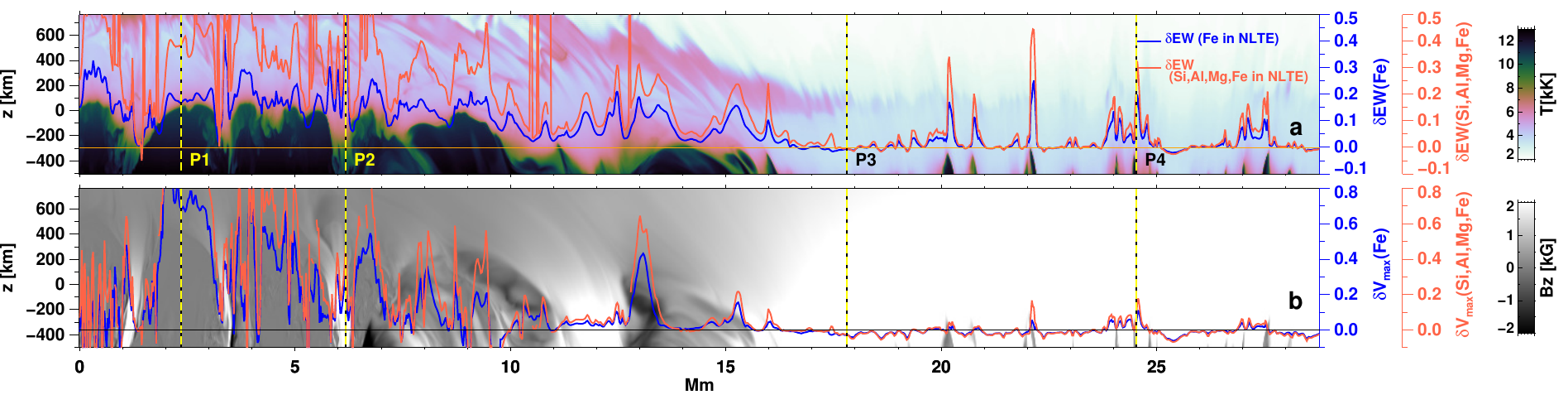}
    \caption{The two panels show the temperature and line-of-sight magnetic field variation over a slice of the atmosphere from where the four pixels were chosen. The pixel positions are marked by vertical lines. {The variation in $\delta EW$ (panel a) and $\delta V_{\rm max}$ (panel b) across this slice are shown in blue and red curves.} The former represents the case where only Fe is in NLTE while in the latter {Si, Al and Mg} are also treated in NLTE. See Sections~\ref{sec:detailed_comp} and \ref{sec:detailed_comp_ed} for more details.}
    \label{fig:stokes_delE_ed}
\end{figure*}

\begin{figure*}[htbp]
    \centering
    \includegraphics[width=0.90\textwidth]{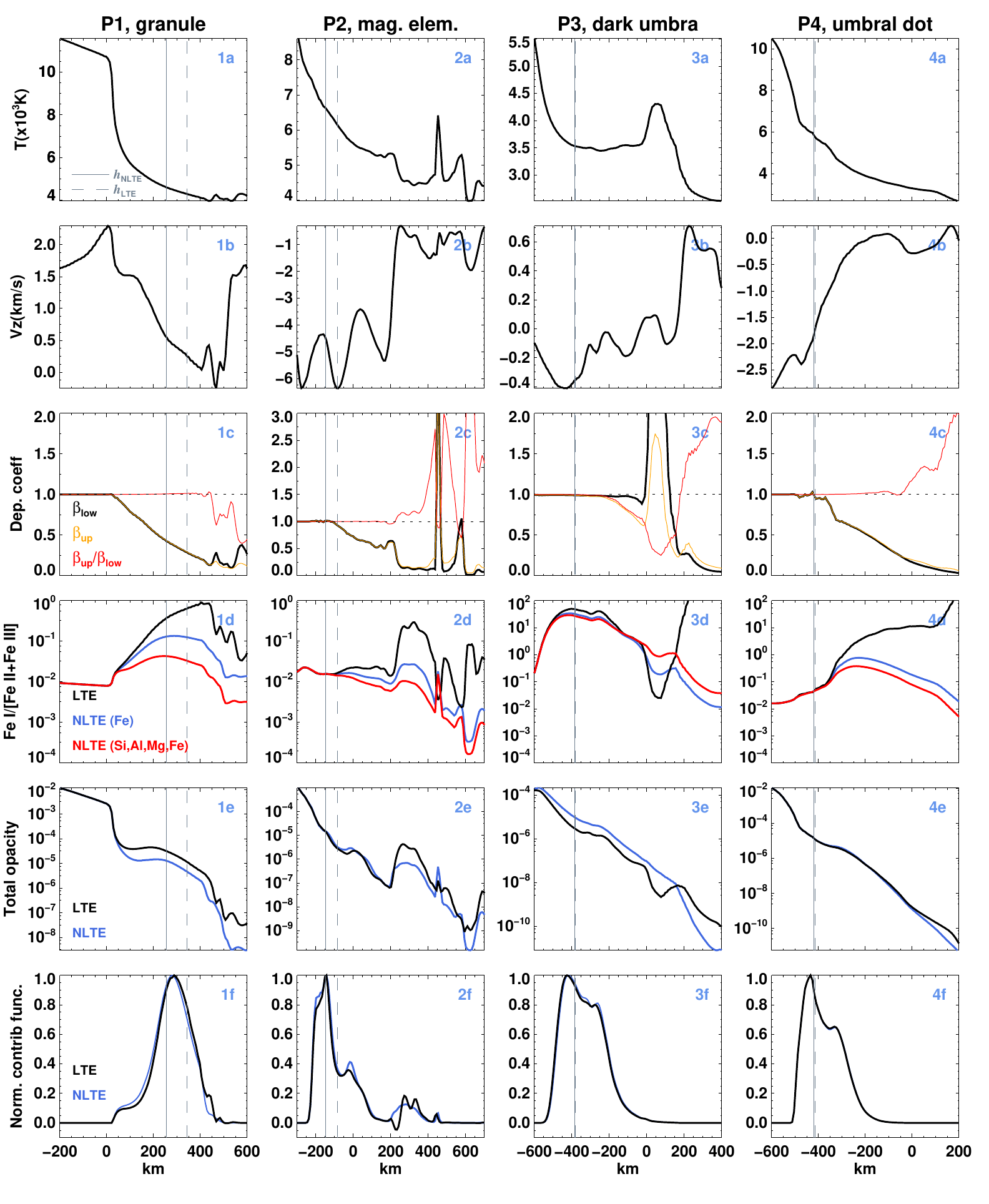}
    \caption{Variation of different physical quantities as a function of height at the four sample pixels. The first two rows show temperature and vertical velocity. In the third and fourth rows, we have plotted the departure coefficients and the ratio of neutral iron atoms to ionized iron atoms (\ion{Fe}{I}/[\ion{Fe}{II} + \ion{Fe}{III}]), respectively for LTE and NLTE computations (distinguishing between the case where only iron is in NLTE - blue - and when the Si, Al and Mg are also treated in NLTE - red). The total opacity at the rest wavelength of the line ($\lambda_0$) is plotted in the fifth row. Finally, {the LTE and NLTE} normalized contribution functions are shown in the bottom row. The vertical lines indicate the height where $\tau_{\lambda_0}=1$ when the line is formed in LTE ($h_{\rm LTE}$, dashed line) or in NLTE ($h_{\rm NLTE}$, solid line) conditions.}
    \label{fig:four_pix_params}
\end{figure*}    

\subsection{Intensity}
\label{subsec:int}
In panel a of Figure~\ref{fig:ew}, the $\delta EW$ is predominantly greater than zero. For the pixels within the orange box, we show a scatter plot of $\delta EW$ as a function of temperature at $z=0$\,km in panel d. Clearly, the relative change in $\delta EW$ becomes larger as the temperature increases. {This is due to the increase in $(J_\nu - B_{\nu})$ excess as the region becomes hotter, a correlation noted by \citet{1992A&A...265..257B, 1994A&A...288..860C}. Due to the higher contrast between hotter and cooler regions in layers close to the surface, this correlation is best seen when we consider the temperature at $z=0$\,km (surface at which the optical depth at 5000\,\AA{} is unity) in panel d of Figure~\ref{fig:ew}, especially over the orange box outside the sunspot.} %

In brighter features such as granules, penumbral filaments, and umbral dots, the $\delta EW$ is found to be as high as $30\%$ (Figure~\ref{fig:ew}, panel a). Also in magnetic structures embedded in the intergranular lanes, the relative change is sometimes higher than $30\%$, indicating the strong influence of NLTE effects in such regions. 
This is also consistent with the findings of \cite{1988A&A...189..243S} where the authors investigated the NLTE effects in both quiet Sun and flux tube models for over 200 iron lines including the one at 6173\,\AA. From their analysis, the authors conclude that the departures from LTE in flux tube models are larger than in the quiet Sun models. In contrast to this, the values of departure coefficients $\beta_{\rm low}$ and $\beta_{\rm up}$ in the magnetic structures seen from Figure~\ref{fig:dep-coeff} are close to one at $h_{\rm NLTE}$. We try to resolve this interesting discrepancy in Section~\ref{sec:detailed_comp}.

In darker features such as the umbra, intergranular lanes, and penumbral spines, the $\delta EW$ is close to zero. This is consistent with the departure coefficients also being close to one in these features (Figure~\ref{fig:dep-coeff}), indicating that the ratios of neutral to ionized iron in such regions are close to their LTE predictions {near the heights of formation of the line. P3 and P4 are examples of two such pixels which will be discussed in detail in \ref{sec:detailed_comp_ed} (see also Figure~\ref{fig:four_pix_params})}. If one observes closely, faint red patches can be seen in the umbra where $\delta EW$ is marginally less than zero. This is due to line strengthening by resonance scattering. Albeit weak, its effects become apparent in the absence of UV overionization. 

In comparison, the line pair at 6300\,\AA{} shows a slightly different behaviour. Its $\delta EW$ maps are presented in Figure~\ref{fig:ew}, panels g and h. The maps contain a mix of grey and red patches {with the grey patches being much lighter compared to those in panel a} for the 6173\,\AA{} line. The LTE departures for the 6300\,\AA{} line pair are less than the departures measured for the 6173\,\AA{} line. The $\delta EW$ is greater than zero in the granules due to the dominance of UV overionization effects. In the intergranular lanes and magnetic structures the $\delta EW$ can be greater or less than zero depending on the extent to which resonance scattering compensates for the UV overionization effects. Similar conclusions can also be drawn from the scatter plot in panel m.

For the 6173\,\AA{} line, the  departures from LTE in the intensity can be clearly seen even when the profiles are spatially averaged (Figure~\ref{fig:ew}, panel 1). Here we average over the orange box since outside it, the NLTE effects are either absent or quite weak. Interestingly, similar spatial averaging of the intensity profiles for the 6301\,\AA{} line, over the same patch, cancels out the NLTE effects from UV overionization and resonance scattering. This is clear from Figure~\ref{fig:ew} panel 4, where the LTE and NLTE profiles are nearly identical. Note that these profiles do not yet include the effects of Si, Al and Mg atoms which set the right photon density in the UV and further enhance the overionization effects (see Section~\ref{sec:detailed_comp_ed}).

\subsection{Polarization}
\label{subsec:pol}
For the 6173\,\AA{} line, the maps of $\delta Q_{\rm max}$ and $\delta V_{\rm max}$ (Figure~\ref{fig:ew}, panels b and c) are similar to the $\delta EW$ maps. That is, the departures from LTE are the largest in the granules. In the intergranular lanes and magnetic structures, we mostly see grey and white patches, and only a few pixels are coloured red. 
Consequently, in a majority of the pixels, the NLTE polarization profiles are weaker than the LTE ones. This is confirmed by the scatter plots in Figure~\ref{fig:ew} (panels e and f) of $\delta Q_{\rm max}$ and $\delta V_{\rm max}$, over the area marked by the orange box, plotted as a function of $B$ at $z=0$\,km. Here again, we see a clear trend. For a majority of the pixels, the relative change in the polarization is positive, indicating the dominance of UV overionization effects. Inverting such profiles in LTE will result in an underestimation of the magnetic field strength. Recently, \citet{2022A&A...661A..59D}, using a 16-level Fe atom model, found that NLTE effects in the 6173\,\AA{} line can introduce differences of up to $60\%$ in the inverted magnetic field strength. 

In the umbra, we again see faint red patches where the amplitudes of the LTE $Q$ and $V$ profiles are larger than in the NLTE case. In the absence of UV overionizaton, weak signatures of resonance scattering start to show up. 

The $\delta Q_{\rm max}$ and $\delta V_{\rm max}$ maps for the 6300\,\AA{} line pair show more red patches in the intergranular lanes in between the grey patches in the granules (Figure~\ref{fig:ew}, panels i - l). These red patches are also stronger and larger. For this line pair, NLTE effects can make the profiles stronger or weaker depending on the atmospheric conditions. Accordingly, the scatter plots in panels n and o, of Figure~\ref{fig:ew}, show a much larger spread in both $\delta Q_{\rm max}$ and $\delta V_{\rm max}$ on either side of zero.   

Similar to the intensity profiles, the spatially averaged polarization profiles of the 6173\,\AA{} line retain the effects of UV overionization, see Figure~\ref{fig:ew} panels 2 and 3. This is particularly striking for Stokes~$Q$. For the 6301\,\AA{} line, the spatial averaging dampens the NLTE effects. Over the region chosen in Figure~\ref{fig:ew}, the NLTE effects are no longer apparent in spatially averaged $V/I_c$ but do survive in $Q/I_c$.  

\begin{figure*}[h!]
    \centering
    \includegraphics[width=0.9\textwidth]{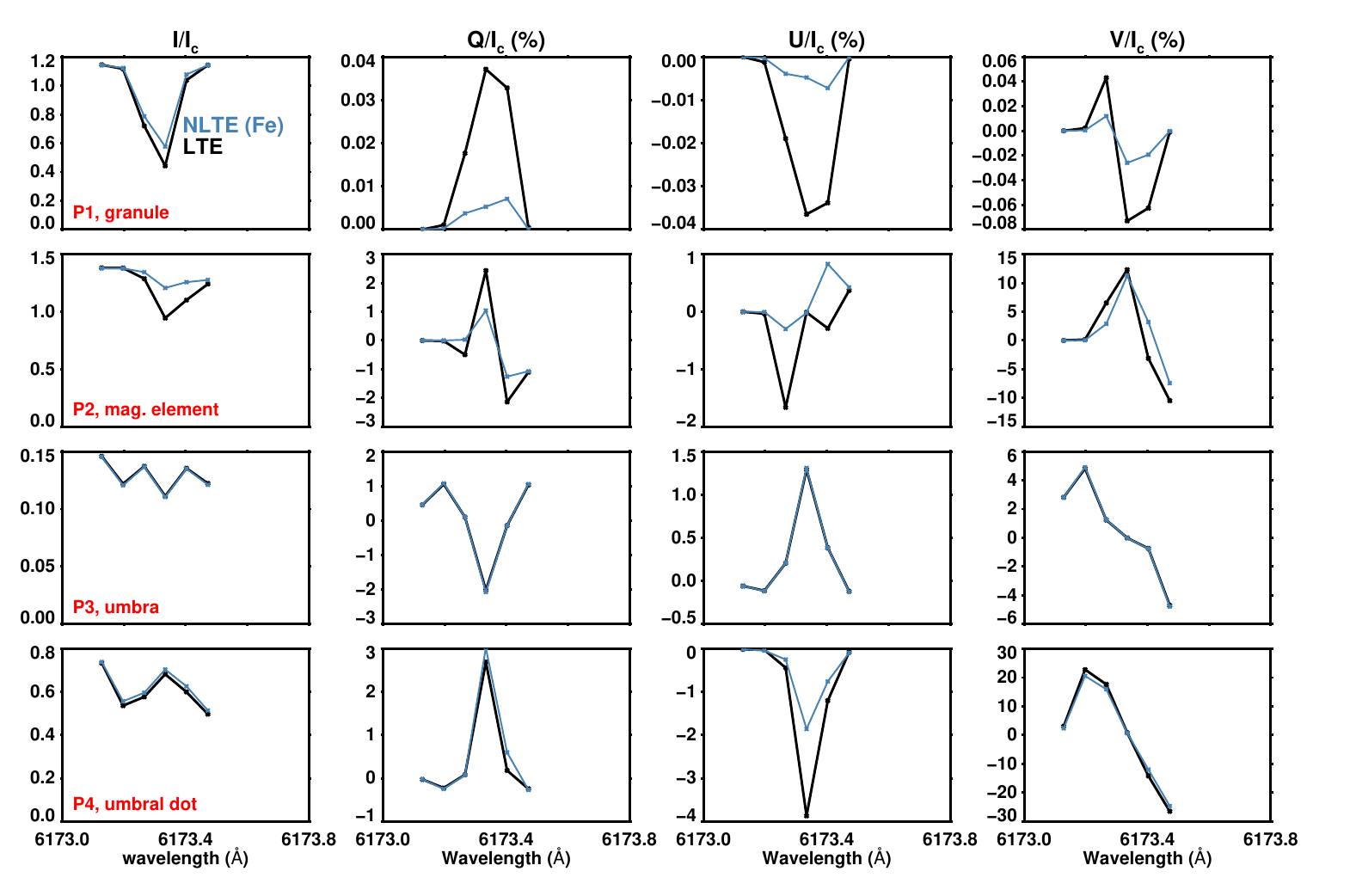}
    \caption{The Stokes profiles at the four sample pixels P1, P2, P3 and P4 after degradation onto a coarse wavelength grid with only six spectral points used by the SDO/HMI instrument. These six points are marked by vertical lines in Figure~\ref{fig:stokes_lte_nlte}. The NLTE profiles used for degradation were computed by treating only Fe in NLTE.}
    \label{fig:stokes_prof_hmi}
\end{figure*}

\subsection{Detailed comparison of the Stokes profiles}
\label{sec:detailed_comp}
We have chosen four sample pixels P1-P4 marked on the continuum intensity map in Figure~\ref{fig:cont} to analyse the NLTE effects in detail. Their Stokes profiles computed in NLTE and in LTE are plotted in Figure~\ref{fig:stokes_lte_nlte}. Pixel P1 is in a granule, P2 is in a small-scale magnetic concentration (or magnetic element), P3 sits in the dark umbra, and P4 is at an umbral dot. In this section we compare the LTE profiles (in black) with the NLTE profiles plotted in blue. These correspond to the case when only iron is treated in NLTE. In the same panels, the red curves were computed by treating also Si, Al and Mg in NLTE, and will be discussed in Section~\ref{sec:detailed_comp_ed}.

In Figure~\ref{fig:stokes_delE_ed},  the panels a and b show temperature and line-of-sight magnetic field variation across the slice from where the four pixels were chosen. The four pixels are marked by vertical lines. The $\delta EW$ for the case with only iron in NLTE is indicated by the blue curve. The $\delta EW$ is close to zero in the umbral region with variations seen at the locations of the umbral dots. Outside the spot, the relative change is greater than zero, in agreement with Figure~\ref{fig:ew}.

The differences in the Stokes profiles are clear at P1 and P2. In P1, the NLTE effects decrease the $EW$ by $17\%$, in P2, by $50\%$. The change in amplitude is close to $70\%$ in P1 for all three polarized Stokes profiles, all of which admittedly have rather small amplitudes.  More interestingly, the amplitudes of the two orders of magnitude stronger $Q/I_c$ and $U/I_c$ profiles in P2 are also reduced by $\approx 70\%$. Here the NLTE Stokes $V/I_c$ profile is not only weakened by $40\%$, but there is also a qualitative change in its shape (from multi-lobed, complex profile to a simple $V$ profile. This is due the strong velocity gradients in that pixel (see panel 2b, Figure~\ref{fig:four_pix_params}).  In P3, the LTE and NLTE profiles nearly match. Finally, in the umbral dot pixel P4, we again see differences, these being most pronounced in the $U/I_{c}$. 

In Figure~\ref{fig:four_pix_params}, we have plotted various quantities such as the temperature, vertical velocity, departure coefficients, ratio of neutral to ionized iron atoms, total opacity and the normalised contribution function, at these four pixels. The heights $h_{\rm NLTE}$ and $h_{\rm LTE}$ are marked by vertical solid and dashed lines respectively. According to this figure, the departure coefficients $\beta_{\rm low}$ and $\beta_{\rm up}$ follow the temperature stratification, as noted earlier by \cite{1992A&A...265..237B}. {On average, the upper level departure coefficient $\beta_{\rm up}$ follows $\beta_{\rm low}$ up to roughly 400 km above the formation height of the local continuum in that pixel}.
Strong gradients and fluctuations in the temperature stratification, typical of high-resolution MHD cubes, makes it challenging to discern a general correlation between the behaviour of the departure coefficients and atmospheric variations, unlike in the one-dimensional atmospheres, such as the VALC model, used in previous studies \citep{1992A&A...265..237B, 1992A&A...265..257B, 1994A&A...288..860C} 

In pixel P1, both $\beta_{\rm low}$ and $\beta_{\rm up}$ are much less than one near the formation heights of the 6173\,\AA{} line. However, their ratio stays close to one at these heights (panel 1c). Here $h_{\rm LTE} - h_{\rm NLTE} \approx 100$\,km. A drop in the ratio of neutral to ionized iron due to UV overionization, compared to their LTE values, can be seen in panel 1d. When the UV line haze is not taken into account, the ratio drops further, enhancing the NLTE effects to unrealistically large values (not shown in the figure). The red curve in panel 1d represents the neutral to ionized iron ratio when Si, Al, and Mg are explicitly treated in NLTE. A detailed discussion on this is presented in Section~\ref{sec:detailed_comp_ed}. The NLTE total opacity at $\lambda_0$, plotted in panel 1e, shows a drop compared to its LTE value around $h_{\rm NLTE}$. Since there is no change in the continuum opacity around 6173\,\AA{}, this drop is due to the change in line opacity. The normalized contribution function is plotted in panel 1f. Although it peaks close to $h_{\rm NLTE}$, the contribution function itself is quite broad,  implying that the line samples a range of heights around $h_{\rm NLTE}$.

Pixel P2 presents quite an interesting case. This pixel is at the edge of a magnetic structure where the temperature and velocity profiles show strong gradients (panels 2a and 2b). The different parameters plotted in panels 2c-2e, are close to their LTE values at $h_{\rm NLTE}$, and yet we see departures from LTE in the Stokes profiles (Figure~\ref{fig:stokes_lte_nlte}). The $h_{\rm LTE} - h_{\rm NLTE}$ is around 50\,km. This can be explained by the multiple peaks in its contribution function (panel 2f). The fragmented nature of this function could arise due to the velocity, which also has similar multiple peaks at those heights (panel 1b). Such steep gradients give rise to a highly asymmetric and distorted intensity profile such as the one plotted in Figure~\ref{fig:stokes_lte_nlte} at P2.  At the locations of the secondary peaks in the contribution function, $\beta_{\rm low} < 1$ and we see departures from LTE in other parameters as well, all of which affect the formation of the line. We find similar multiple peaked contribution functions in and around other magnetic structures. This explains why the $\delta EW$ in Figure~\ref{fig:ew} is greater than zero in magnetic regions (Figure~\ref{fig:stokes_lte_nlte}, panels a and b) despite the departure coefficients $\beta_{\rm low}$ and $\beta_{\rm up}$ being close to unity at such locations in Figure~\ref{fig:dep-coeff}.

Because spectral lines are formed over a range of heights, the physical conditions at the reference height, $h_{\rm NLTE}$, alone provide an incomplete picture on what influences the formation of a spectral line.  Defining a height of formation or looking at the atmospheric conditions at a single height could be misleading \citep{1996A&A...314..295S}. But making use of reference heights such as $h_{\rm NLTE}$ and $h_{\rm LTE}$, helps us compare different quantities to explain the complex NLTE mechanisms. It gives us a rough idea about the different atmospheric layers sampled by the spectral line. In addition, the wavelength chosen to define $\tau_{\lambda}=1$ introduces further uncertainty. Here we have chosen $\lambda=\lambda_0$, the rest wavelength of the line. But the actual line core wavelength need not coincide with the rest wavelength of the line. 

In the umbra, we find that the effect of UV overionization is negligible. The UV photons are much less superthermal over such cooler regions \citep{1992A&A...265..257B}. {At pixel P3 chosen from the umbra, the departure coefficients are close to unity and the ratio of neutral to ionized iron is close to its LTE value around the height of formation of the line (see panels 3c and 3d in Figure~\ref{fig:four_pix_params}).} The striking thing about the umbra is that it is the only place where within the range of line formation \ion{Fe}{I} is the dominant ionisation stage of iron, see panel 3d of Figure~\ref{fig:four_pix_params}. This also partly explains why there is no overionisation there. Since there is so much \ion{Fe}{I}, changing the ionisation rate by a factor of two still leaves it as the dominant species and largely unchanged, while the number of atoms of the minority species, \ion{Fe}{II} in this case, changes by nearly a factor of 2. 

Another interesting feature in the umbra is that in panel 3e of Figure~\ref{fig:four_pix_params}, the NLTE total opacity is slightly larger than LTE opacity up to 180\,km. As previously discussed in Section~\ref{subsec:int}, this is due to resonance scattering. Its effects become apparent in the absence of UV overionization. For the 6173\,\AA{} line, this is quite weak. From Figure~\ref{fig:stokes_lte_nlte}, the LTE and NLTE profiles nearly match. 

The umbral dots are brighter than their surrounding dark umbra and we see signatures of UV overionization in them although not as strongly as in the granules. At $h_{\rm NLTE}$, the parameters in panels 4c - 4e do not show much deviation from their LTE values. The broad contribution function implies that the line is sampling also layers higher than $h_{\rm NLTE}$ and at these heights the NLTE parameters deviate from their LTE values. This explains the differences in Stokes profiles observed in Figure~\ref{fig:stokes_lte_nlte}.

\subsection{Importance of treating Si, Al and Mg in NLTE}
\label{sec:detailed_comp_ed}
{The amount of excess photoionization of the iron atoms is set by the radiation field in the UV. For a proper NLTE treatment of the iron line, it is therefore crucial to compute the correct photon density in the UV.  Metals Si, Al, and Mg all have ionization edges in the UV and they have a strong influence on $J_\nu$ which  in turn affects the ionization equilibrium of iron. In addition, these metals are also the most important electron donors in the temperature minimum region, where hydrogen is not fully ionized \citep{1981ApJS...45..635V}. A full NLTE computations of the iron lines will then involve a NLTE treatment of the metals Fe, Si, Mg, and Al along with electron densities being computed in NLTE. The latter is beyond the scope of this present study. To ensure that we use the correct levels of $J_\nu$ in the UV, we add the contributions from Si, Al, and Mg by treating them in NLTE while setting the electron densities to their LTE values.}

Over the stripe indicated by the yellow line in the continuum image shown in Figure~\ref{fig:cont}, we computed the Stokes profiles by treating \ion{Si}{I}, \ion{Mg}{I} and \ion{Al}{I} in NLTE in addition to \ion{Fe}{I}. For \ion{Si}{}, we used a 16 level model atom (15 \ion{Si}{I} levels + ground state of \ion{Si}{II}) coupled by 28 line transitions and 15 continuum transitions. We included \ion{Mg}{} using a model atom with 11 \ion{Mg}{I} levels plus the ground state of \ion{Mg}{II}. The model atom includes 15 bound-bound and 11 bound-free transitions. For \ion{Al}, we use an 18 level model atom of which 9 levels belong to \ion{Al}{I}, 8 levels to \ion{Al}{II} and the last level is the ground state of \ion{Al}{III}. These levels are coupled by 19 bound-bound and 17 bound-free transitions.

The Stokes profiles computed by treating Si, Al and Mg atoms in NLTE at the four sample pixels are shown by the red curves in Figure~\ref{fig:stokes_lte_nlte}. The $\delta EW$ over the whole stripe are plotted in the bottom two panels of Figure~\ref{fig:stokes_delE_ed}. Clearly the NLTE effects are much stronger now compared to the case when only Fe was in NLTE. The intensity profiles are shallower and the $\delta EW$ values are larger, as shown in panel a of Figure~\ref{fig:stokes_delE_ed}. {Similarly, the polarization profiles are also affected. Inclusion of Si, Al, and Mg in NLTE further weakens the Stokes $V$ amplitude. The trend over the whole stripe can be seen in panel b of Figure~\ref{fig:stokes_delE_ed}.} 
The iron overionization and the line opacity deficit has increased. This is also evident from the reduced ratio of neutral to ionized iron at the four sample pixels plotted in Figure~\ref{fig:four_pix_params} (red curves, panels 1d, 2d, 3d and 4d).  

In pixel P1, the relative change in polarization amplitudes with respect to the LTE values are now as high as $90\%$. In the four Stokes profiles, the inclusion of {Si, Al and Mg in NLTE} has increased the LTE departures by nearly $ 20\%$ compared to the case when only iron was treated in NLTE. 
 In P2, the $\delta EW$ is now $61\%$ which is a $10\%$ increase. In the polarization profiles, $\delta Q_{\rm max}$, $\delta U_{\rm max}$ and $\delta V_{\rm max}$ increase between $5\%$ and $10\%$. In P3 and P4, the enhancement in relative change is quite small except in $U/I_c$ profile at P4, where we see a nearly $25\%$ increase.

Thus for a better treatment of NLTE, it is important to treat also Si, Al and Mg in NLTE, in addition to iron. Their impact on the Stokes profiles is once again dependent on the atmosphere.

\subsection{Effects of spectral degradation}
\label{subsec:stokes_degrad}
The observations in the \ion{Fe}{I} 6173\,\AA{} line from both ground-based and space-based telescopes are at a much lower spectral resolution than what is used by us in this study. To investigate the effects of spectral degradation, we choose the coarse wavelength grid used by the SDO/HMI instrument. The effects of different filter curves were approximated by convolving with a single Gaussian of FWHM 75\,m\AA, and interpolating the profiles onto the six wavelength points sampled by HMI (neglecting shifts introduced by the relative spacecraft-Sun velocity). This was done for both LTE and NLTE (Fe only) profiles. Their comparison is presented in Figure~\ref{fig:stokes_prof_hmi}. Both, the degradation in the profile shape due to coarse wavelength sampling and the large effects of departure from LTE are clearly seen in the figure. Comparing the profiles in Figure~\ref{fig:stokes_lte_nlte} and \ref{fig:stokes_prof_hmi} highlights that when the spectrum is sampled with such a coarse grid, the poor spectral sampling of the line has a bigger effect on the Stokes profiles than the NLTE effects. Nevertheless, the fact that NLTE effects survive spectral degradation indicates that they will likely survive in observations from other telescopes such as SO/PHI or Vigil/PMI, and also in SST and DKIST which can observe this line at higher spectral resolutions. Analysis or inversions of such observations assuming LTE will introduce errors in the inverted atmosphere. 

\section{Conclusions}  
\label{sec:conclusions}
In this paper, we have for the first time presented a detailed analysis of the NLTE formation of the widely used photospheric iron line at 6173\,\AA, by synthesizing its Stokes profiles in both LTE and NLTE using a 3D-MHD cube. Understanding its formation in an MHD cube is quite complex. In particular, the strong gradients in the cube make it challenging to correlate the LTE departures seen in the spectra to the nature of NLTE effects playing a role. 

The line 6173\,\AA{} is an LTE line considering its source function. That is, resonance scattering in the line is quite weak due to its low excitation potential and line interlocking effects. However, the NLTE effects acting on the iron species as a whole, namely the UV overionization, plays an important role in shaping its Stokes profiles. 

In comparison, the iron line pair at  6301\,\AA-6302\,\AA{} are not only affected by UV overionization but also by scattering  \citep{2001ApJ...550..970S, 2021arXiv210302369R}. The UV overionization introduces a line opacity deficit while scattering results in a line source function deficit. They affect the line profiles in opposite ways, reducing the net deviation from LTE profiles. 

The departures from LTE for the 6173\,\AA{} line are stronger compared to the 6301\,\AA-6302\,\AA{} line pair. For the 6173\,\AA{} line, the NLTE intensity and polarization profiles tend to be weaker than the LTE profiles.  This is particularly clearly seen in bright features such as granules and bright magnetic structures, but to some extent also in penumbral filaments and umbral dots. In darker regions such as intergranular lanes, dark umbra, penumbral spines, the NLTE profiles match closely with the LTE ones.

In order to reproduce the right amount of UV overionization in the Fe lines, it is necessary to treat other prominent metals such as Si, Mg and Al in NLTE as well. This is important to set the right ionization equilibrium for iron through $J_\nu$. 
We find that this increases the departure from LTE as compared to the case where only Fe is treated in NLTE. Table~9 and Figure~36 of \citet{1981ApJS...45..635V}, suggest that \ion{Si}{I} and \ion{Mg}{I} are the most important in setting the right $J_\nu$ in the UV \citep[see also][]{2021arXiv210302369R}. Hence it is important to treat at least these two species in NLTE along with Fe to properly account for the UV overionization effects in the atmosphere. {These metals are also dominant electron donors in the temperature minimum. In the present paper, however, we do not take this into account and compute the electron densities in LTE. Consequently, our NLTE treatment is still incomplete. For chromospheric lines such as \ion{Mg}{II} k, \citet{2019A&A...631A..33B} showed that the differences in electron densities between LTE and NLTE computations have an impact on the intensity profiles. For the photospheric iron lines such as the \ion{Fe}{I} 6173\,\AA{}, this remains to be investigated.}

The NLTE effects in the 6173\,\AA{} line continue to survive even when the Stokes profiles are spatially averaged or spectrally degraded. In the present paper, this was tested only on profiles computed by treating the iron atom alone in NLTE. {Since the treatment of Si, Al and Mg in NLTE enhances the LTE departure in spatially resolved profiles, it will likely have a similar effect on the spatially averaged profiles as well.} This means, when a complete treatment of NLTE is carried out over the whole cube then the spatially averaged NLTE profile will likely show a larger deviation from LTE compared to the case presented in Figure~\ref{fig:ew} (panels 1-3).

 Modeling the observed iron profiles using LTE inversion codes will compensate for the above NLTE effects by tweaking the temperature and other atmospheric parameters, an effect known as NLTE-masking \citep{1982A&A...115..104R, 1988ASSL..138..185R}.  In the case of 6173\,\AA{} line, since the NLTE effects act in a unidirectional manner, its LTE inversions are expected to result in an overestimation of the temperature and an underestimation of the magnetic field strength. 

\begin{acknowledgements}
This work has greatly benefited from the discussions with R. J. Rutten. HNS would also like to thank L. P. Chitta and D. Przybylski for their valuable inputs.
\end{acknowledgements}

\end{document}